\documentclass[11pt]{article}
\usepackage{margins,epsfig,amstex,xspace,amssymb}

\newcommand{\pct}{\%\xspace}
\newcommand{\eps}{\ensuremath{\epsilon}}
\newcommand{\lrar}{\ensuremath{\leftrightarrow}}
\newcommand{\EL}[2][]{\ensuremath{^{#1}\rm{#2}}}
\newcommand{\me}{\ensuremath{\left| \mathcal{M} \right|^2}}
\newcommand{\he}{\ensuremath{^4\rm{He}}\xspace}

\begin{document}
\baselineskip=16pt
\pagestyle{empty}
\begin{center}
\rightline{Fermilab--Pub--98/232A}
\rightline{astro-ph/9807279}
\rightline{submitted to {\em Physical Review D}}
\vspace{.2in}

\bibliographystyle{aip}

{\bf Precision Prediction for the Big-Bang Abundance of Primordial $^4$He}\\
\vspace{.2in}

Robert E. Lopez$^{1,2}$ and Michael S. Turner$^{1,2,3}$\\
\vspace{.1in}

$^1${\it Department of Physics,\\
The University of Chicago, Chicago, IL  60637-1433}\\
$^2${\it NASA/Fermilab Astrophysics Center, \\
Fermi National Accelerator Laboratory, Batavia, IL  60510-0500}\\
$^3${\it  Department of Astronomy \& Astrophysics,\\
Enrico Fermi Institute,
The University of Chicago, Chicago, IL 60637-1433}\\
\end{center}

\vspace{.3in}

\centerline{\bf ABSTRACT}
\bigskip

\noindent 
Within the standard models of particle physics and cosmology we have calculated the
big-bang prediction for the primordial abundance of \he to a theoretical uncertainty
of less than $0.1\,\pct$ $(\delta Y_P < \pm 0.0002)$, improving the current
theoretical precision by a factor of 10. At this accuracy the uncertainty in the
abundance is dominated by the experimental uncertainty in the neutron mean lifetime,
$\tau_n = 885.4 \pm 2.0\,\rm{sec}$. The following physical effects were included in
the calculation: the zero and finite-temperature radiative, Coulomb and
finite-nucleon-mass corrections to the weak rates; order-$\alpha$
quantum-electrodynamic correction to the plasma density, electron mass, and neutrino
temperature; and incomplete neutrino decoupling. New results for the
finite-temperature radiative correction and the QED plasma correction were used. In
addition, we wrote a new and independent nucleosynthesis code designed to control
numerical errors to be less than 0.1\pct. Our predictions for the \EL[4]{He} abundance
are presented in the form of an accurate fitting formula.  Summarizing our work in
one number, $ Y_P(\eta = 5\times 10^{-10}) = 0.2462 \pm 0.0004\ ({\rm expt})\ \ \pm <
0.0002\ ({\rm theory})$. Further, the baryon density inferred from the Burles-Tytler
determination of the primordial D abundance, $\Omega_B\,h^2 = 0.019\pm 0.001$, leads
to the prediction: $Y_P = 0.2464 \pm 0.0005\,\rm{(D/H)}\,\pm < 0.0002
\,\rm{(theory)}\, \pm 0.0005\,\rm{(expt)}$. This ``prediction'' and an accurate
measurement of the primeval \he abundance will allow an important consistency test of
primordial nucleosynthesis.

\newpage
\pagestyle{plain}
\setcounter{page}{1}

\section{Introduction}

Big-bang nucleosynthesis (BBN) is one of the observational pillars of the standard
cosmology. Further, it has the potential to be a precision probe of the early
universe and fundamental physics~\cite{Schramm98,Yang84,Walker91}. Observations of
light-element abundances have improved dramatically over the past few years, and the
current and planned precision measurements of D, \he, $^3$He and $^7$Li, should allow
a precise (10\pct or better) determination of the baryon density and consistency
check of BBN, but only if the theoretical predictions of the light element abundances
are as good as the observations. In particular, a measurement of the primeval D
abundance pins down the baryon density, and in turn makes predictions for the other
three abundances. Because the subsequent evolution of the $^4$He abundance is simple
- stars make \he - and measurements have the potential of determining $Y_P$ to three
significant
figures~\cite{Izotov98,Izotov982,Olive95,Olive97,Olive972,Pagel92,Pagel922}, \he can
provide an important consistency check of BBN. Furthermore, an independent
determination of the baryon density from cosmic microwave background anisotropies
will soon test the consistency of the standard model of cosmology. Finally, the
combination of accurate observations and theory can be used to test physics beyond
the standard model of particle physics~\cite{Schramm98,Sarkar96}, e.g., by imposing a
strict limit on the number of light neutrino
species~\cite{Copi97,Schvartsman69,Steigman77}. Cosmology is entering a high
precision age, and this motivates high-precision BBN predictions.

Over the years, theoretical study of \he synthesis has been intense, with the
following effects being considered: Coulomb and radiative corrections to the weak
rates~\cite{Dicus82,Donoghue83,Donoghue84,Kernan93,Sawyer96,Chapman97}, BBN code
numerical errors~\cite{Kernan93}, nuclear reaction rate
uncertainties~\cite{Smith93,Fiorentini98}, finite-temperature QED plasma
corrections~\cite{Dicus82,Heckler94}, the effect of finite-nucleon
mass~\cite{Seckel93,Gyuk93,Lopez97}, and incomplete neutrino
decoupling~\cite{Dicus82,Dodelson92}. However, the corrections have been incorporated
in a patchwork fashion and a recent informal poll of BBN codes indicated a spread of
1 \pct in the predicted value of the \he abundance for fixed $\eta$ and $\tau_n$.

The goal of this work was a calculation of the primordial abundance of \he, within
the standard models of particle physics and cosmology, accurate enough so that its
uncertainty is dominated by the experimental uncertainty in the neutron mean
lifetime\footnote{The Particle Data Group currently recommends $\tau_n = 887\pm 2$
sec~\cite{PDG98}. A recent measurement using ultracold neutrons indicates a slightly
lower value, $\tau_n = 885.4 \pm 0.9 \,\rm{(stat)} \pm 0.4
\,\rm{(sys)}\,\rm{sec}$~\cite{Arzumanov97}. For our central value we use 885.4 sec and
for the uncertainty we use $\pm 2$ sec.}, $\tau_n = 885.4 \pm
2.0\,\rm{sec}$~\cite{PDG98,Arzumanov97,Byrne96}. Because $\tau_n$ is so accurately
known ($\delta\tau_n/\tau_n = 0.23\pct$), it is used to normalize all of the weak
rates that interconvert neutrons and protons: $ep
\leftrightarrow \nu n$, $e^+n
\leftrightarrow
\bar{\nu}p$ and $n \leftrightarrow pe\bar{\nu}$. The baryon-number fraction of $^4$He 
produced ($\equiv Y_P$) depends sensitively on the weak rates because they determine
the neutron-to-proton ratio $n/p$ before nucleosynthesis, and essentially all of the
neutrons around at the onset of nucleosynthesis go into forming $^4$He. We have
determined the effect on $Y_P$ by perturbing the weak rates in the standard
code~\cite{Kawano92},
\begin{equation}
\label{eq:dY_dG}
{\delta Y_P \over Y_P} \simeq - 0.8 {\delta \Gamma \over \Gamma} .
\end{equation}
Since the weak rates scale as $1 / \tau_n$, this estimate implies that $\delta
\tau_n$ introduces an uncertainty in $Y_P$ of 0.18\pct. We use this
uncertainty to set our goal for all theoretical uncertainty. 

To meet our goal we need to calculate the weak rates to precision of better than
0.23\pct. Another source of errors in $Y_P$ come from thermodynamics, i.e., the
energy density $\rho$, the pressure $P$ and the neutrino temperature $T_\nu$. To
determine how accurately we need to know thermodynamic quantities, we can estimate
the change in $Y_P$ due to a change in a thermodynamic quantity, e.g., $\rho$. Again,
using the standard code, we find
\begin{equation}
\label{eq:dY_drho}
{\delta Y_P \over Y_P} \simeq  0.4 {\delta \rho \over \rho} ,
\end{equation}
This indicates that we should calculate thermodynamic quantities to better than
0.45\pct.

When calculating $Y_P$ to this precision, several factors must be considered:
\begin{enumerate}
\item Weak rate and thermodynamics numerics: most quantities to be calculated involve
integrations that must be done numerically. \label{it:ratenum}
\item ODE integration numerics: nucleosynthesis codes contain finite stepsize
errors. \label{it:bbnnum}
\item Nuclear reaction rates: errors originate from experimental uncertainties in the
nuclear reaction data, as well as from neglecting nuclear reactions important to
BBN. \label{it:nuclear}
\item Weak-rate physics: there are several small physical effects that must be
calculated, including Coulomb, zero and finite-temperature radiative corrections, and
the effect of finite-nucleon mass. \label{it:wkphys}
\item Thermodynamics physics: for temperatures much greater than the electron mass,
there are order-$\alpha$ quantum electrodynamic corrections to the equation of state
of the plasma. \label{it:thphys}
\item Incomplete neutrino decoupling: neutrinos share partially in the entropy
release when $e^\pm$ pairs annihilate. \label{it:ind}
\end{enumerate}
Items~\ref{it:ratenum},~\ref{it:bbnnum}~and~\ref{it:nuclear} are addressed in the
next Section; item~\ref{it:wkphys} is addressed in Sec. 3. Items
~\ref{it:thphys}~and~\ref{it:ind} are taken up in Sec. 4, and a summary of our
results is given in the final section.

We mention that we have not considered the ${\cal O}(\alpha^{3/2})$ collective plasma
effects due to the presence of the copious numbers of $e^\pm$ pairs at the time of
BBN, because they are safely below our theoretical error budget of $0.1\,\pct$ for
$Y_P$.  These effects, all of relative size $0.1\,\pct$ and calculated in
Ref.~\cite{Itoh97}, are: the enhancement of nuclear reaction rates due to Debye
screening of nuclear charge; the contribution of longitudinal plasmon modes
($k\lesssim \omega_p \sim 4\pi n_{e^\pm}/T$) to the energy density and pressure; the
(negative) contribution to the energy density and pressure of the electromagnetic
interaction of $e^\pm$ pairs; and the reduction of the energy and pressure of photons
due to plasma effects on low frequency photons ($k\lesssim \omega_p$). Finally, while
we have tried to be exhaustive and very careful in our analysis, we cannot rule out
systematic theoretical error: that is, the possibility that we have neglected some
microphysical effect as important as those we have included.

\section{Numerics}

\subsection{BBN Code}
We have written a new nucleosynthesis code that is independent of the standard
(Kawano) code~\cite{Kawano92}. The heart of any nucleosynthesis code is the set of
ordinary differential equations that govern the evolution of the abundances of the
light elements (see, e.g., Ref.~\cite{Wagoner66,Wagoner73}). Our code tracks protons,
neutrons, D, T, \EL[3]{He}, \EL[4]{He}, \EL[6]{Li}, \EL[7]{Li} and
\EL[7]{Be}. The baryon-number fraction of element $i$ is given
by\footnote{Baryon-number fraction and baryon-mass fraction differ by order 1\pct due
to nuclear binding energy. Because nuclear reactions change the total mass in
baryons, the mass fraction of species $A_i$ ($\equiv X^{mass}_i$) can change even if
the number of species $A_i$ does not. The mass fraction of species $A_i$ is
\[
X^{mass}_i = {n_i m_i \over \sum_j n_j m_j} = 
{n_i\over n_H} \, {m_i\over m_H} \, {1\over 1+\sum_j (n_j/n_H) \, (m_j/m_H)} \, ,
\]
where $m_i$ is the mass of species $i$: e.g., $m_4 = 4.002602\,\rm{amu}$ and $m_H =
1.00783\,\rm{amu}$. For $Y_P=0.25$ and the primordial mix of elements $X^{mass}_4 =
0.24866$. Similarly, the relationship between the baryon mass density and $\eta$
depends on elemental composition. For the primordial mix with $Y_P = 0.25$,
\[
\Omega_B\,h^2 = 3.66 \times 10^7 \, \eta \, ,
\]
with $T_\gamma = 2.7277\,K$. Assuming a mass of 1 amu per nucleon, the prefactor is
$3.639\times 10^7$, and for solar abundance, the prefactor is $3.66043\times 10^7$.}

\begin{equation}
X_i = {A_i n_i \over n_B}
= {A_i\,(n_i/n_H) \over 1+ \sum_i\,A_i\,(n_i/n_H)} \,,
\end{equation}
where $A_i$ is the element's atomic number, $n_i$ it's number density, and $n_B$ is
the baryon-number density. (Note, by convention $Y_P$ is used to denote
$X_4$). Nuclear reaction rates govern the evolution of the elemental
abundances. Conservation of baryon number provides the constraint:
\begin{equation}
\label{eq:cons_mass}
\sum_i X_i = 1.000 \,.
\end{equation}

\begin{figure}
\centerline{\psfig{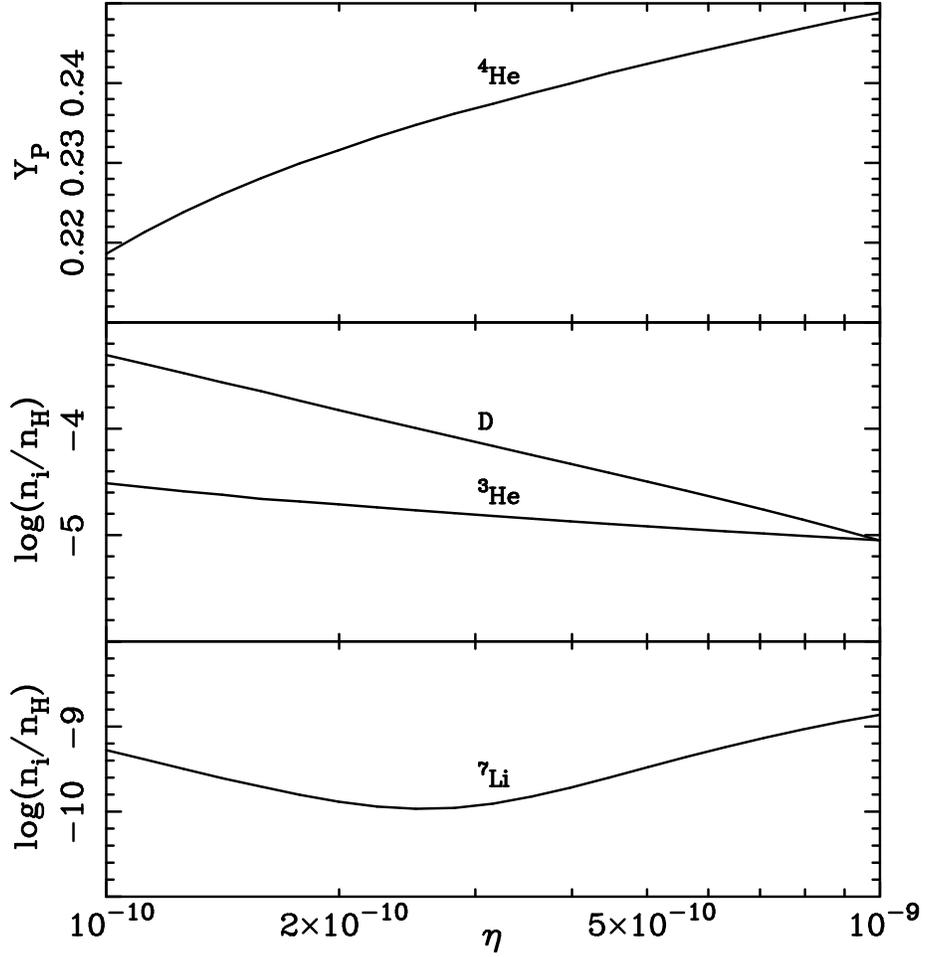}}
\caption{Baseline predictions: element abundances predicted by our BBN
code.}
\label{fg:bbn}
\end{figure}

We take for our initial temperature, $T_i = 10\;\rm{MeV}$, and for our initial
abundances, the nuclear statistical equilibrium (NSE) values:
\begin{equation}
X_{A} = g_A \left[\zeta(3)^{A-1} \pi^{(1-A)/2} 2^{(3A-5)/2} \right] A^{5/2}
	\left({T\over m_N}\right)^{3(A-1)/2} \eta^{A-1} X_p^Z X_n^{A-Z} e^{B_A/T}
\end{equation}
where $A$ is the atomic number, $m_N \simeq 940\,\rm{MeV}$ is the nuclear mass,
$\eta$ is the baryon-to-photon ratio, $B_A$ is the binding energy of species $A$, and
$\zeta(3) \simeq 1.20206$. At temperatures greater than about 1 MeV, the nuclear
rates are sufficiently high to cause the abundances to rapidly assume their NSE
values. ( As discussed in Ref~\cite{Lara98}, the final abundances are very
insensitive to the assumed initial abundances). If we make the well-justified
assumption that the elements are always in kinetic equilibrium, then the rate
coefficients depend only on $\eta$ and $T$. This implies the important and well known
conclusion that the predictions of nucleosynthesis are a function of only one
parameter, $\eta$, which is equivalent to $n_B$ since $T_{\gamma} = 2.7277 \pm 0.002$
K is so well known.

Several important quantities enter into the evolution equations: weak rates,
thermodynamic quantities and nuclear reaction rates. For the weak rates, we define
the total conversion rates (per neutron or proton):
\begin{eqnarray}
\Gamma_{n \rightarrow p} & \equiv & 
\Gamma_{e^+ \: n \rightarrow \bar{\nu} \: p} \: + \:
\Gamma_{\nu \: n \rightarrow e \: p \:} + \: 
\Gamma_{n \rightarrow p \: e \: \bar{\nu}} \,, \nonumber \\
\Gamma_{p \rightarrow n} & \equiv & 
\Gamma_{e \: p \rightarrow  \nu \: n} \: + \:
\Gamma_{\bar{\nu} \: p \rightarrow e^+ \: n} \: + \: 
\Gamma_{p \: e \: \bar{\nu} \rightarrow n} \,.
\end{eqnarray}
Simple expressions for these rates may be obtained assuming no radiative corrections
and infinite nucleon mass. The thermodynamic quantities that must be calculated are
$\rho(T)$, $T_\nu(T)$, $\rho_B(T)$ and the differential time-temperature relation
$dt/dT$.

Our BBN code is independent from the standard code, with one exception: It uses the
same nuclear-rate data (with the exception of the weak rates). The nuclear-reaction
network corresponds to the smallest one offered by the standard code, which contains
the reactions listed in Table~\ref{tb:rates}. Although this network is much smaller
than the largest offered in the standard code, we have verified that the effect on
$Y_P$ of neglecting these additional reactions is less than $10^{-4}$. The
light-element abundances predicted by our code are shown in Fig.~\ref{fg:bbn}.

\begin{table}
\begin{center}
\( \begin{tabular*}{65mm}{r@{{)}{\extracolsep{\fill}}}%
	r@{\hspace{10pt}{\lrar}}l@{}}
1 & \EL{p} + \EL{n} &  \EL{D} + \EL{\gamma}  \\
2 & \EL{D} + \EL{n} & \EL{T} + \EL{\gamma}  \\
3 & \EL[3]{He} + \EL{n} & \EL[4]{He} + \EL{\gamma}  \\
4 & \EL[6]{Li} + \EL{n} & \EL[7]{Li} + \EL{\gamma}  \\
5 & \EL[3]{He} + \EL{n} & \EL{T} + \EL{p}  \\
6 & \EL[7]{Be} + \EL{n} & \EL[7]{Li} + \EL{p}  \\
7 & \EL[7]{Li} + \EL{n} & \EL[3]{He} + \EL[4]{He}  \\
8 & \EL[7]{Be} + \EL{n} & \EL[4]{He} + \EL[4]{He}  \\
9 & \EL{D} + \EL{p} & \EL[3]{He} + \EL{\gamma}  \\
10 & \EL{T} + \EL{p} & \EL[4]{He} + \EL{\gamma}  \\
11 & \EL[6]{Li} + \EL{p} & \EL[7]{Be} + \EL{\gamma}  \\
12 & \EL[7]{Li} + \EL{p} & \EL[4]{He} + \EL[4]{He}  \\
13 & \EL{D} + \EL[4]{He} & \EL[6]{Li} + \EL{\gamma}  \\
14 & \EL{T} + \EL[4]{He} & \EL[7]{Li} + \EL{\gamma}  \\
15 & \EL[3]{He} + \EL[4]{He} & \EL[7]{Be} + \EL{\gamma}  \\
16 & \EL{D} + \EL{D} & \EL[3]{He} + \EL{n}  \\
17 & \EL{D} + \EL{D} & \EL{T} + \EL{p}  \\
18 & \EL{D} + \EL{T} & \EL[4]{He} + \EL{p}  \\
19 & \EL{D} + \EL[3]{He} & \EL[4]{He} + \EL{n}  \\
20 & \EL[3]{He} + \EL[3]{He} & \EL[4]{He} + \EL{p} + \EL{p}  \\
21 & \EL{D} + \EL[7]{Li} & \EL[4]{He} + \EL[4]{He} + \EL{n}  \\
22 & \EL{D} + \EL[7]{Be} & \EL[4]{He} + \EL[4]{He} + \EL{p} 
\end{tabular*} \)
\end{center}
\caption{Reactions used in our code.}
\label{tb:rates}
\end{table}

\subsection{Numerical Accuracy of the BBN Codes}
Because the differential equations governing the light-element abundances are stiff,
an implicit integrator was used to evolve them. Instead of specifying explicit time
steps, as in the standard code, the desired final accuracies are specified as
parameters of our code's integrator. The temperature steps are then determined
adaptively. Integrator accuracy parameters are chosen to be small enough so that
stepsize errors were much smaller than the allowed error in $Y_P$.

\begin{figure}
\centerline{\epsfig{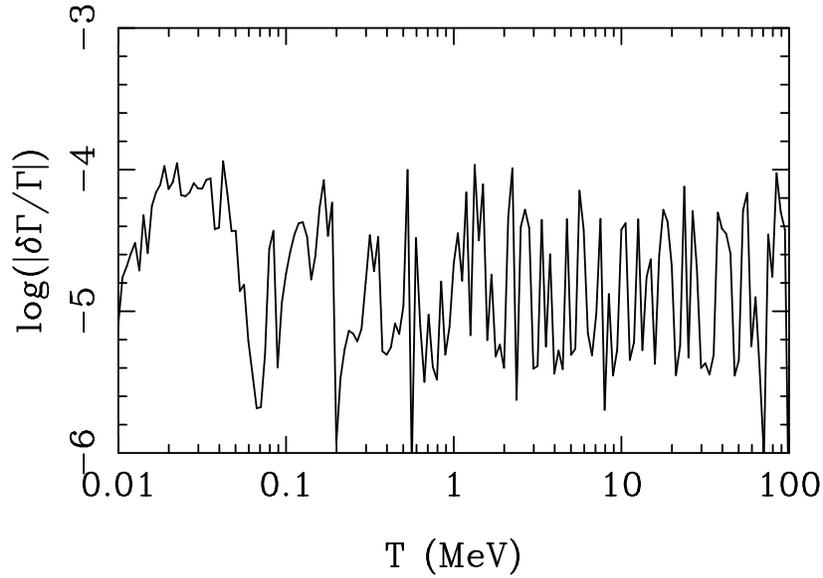}}
\caption{Actual numerical error in calculating $\Gamma_{ep\rightarrow\nu n}$ for error
parameter set at $\delta\Gamma / \Gamma = 10^{-4}$. The error is smaller than the
specified accuracy ($10^{-4}$) for all temperatures. Similar results were obtained
for the other weak rates and thermodynamic quantities.}
\label{fg:numacc}
\end{figure}

To calculate the weak rates and thermodynamic quantities accurately, we proceed as
follows (see, e.g., Ref.~\cite{numrec_rkint}). Let $I = \int_a^b\,f(x)\,dx$ for some
function $f(x)$. Expressed as a first order ordinary differential equation, $I =
J(b)$ where $dJ/dx = f(x)$, $J(a) = 0$. We solve this differential equation using a
fourth order Runge-Kutta routine. Figure~\ref{fg:numacc} demonstrates for a specific
example that the actual numerical errors are as small as requested. All of the weak
rates and thermodynamic quantities were calculated so that their numerical error
contributions to the uncertainty in $Y_P$ were acceptably small.

We compared the results of our code to the standard code, which dates back to the
original version written in 1966~\cite{Wagoner66}, was updated by Wagoner in
1973~\cite{Wagoner73,Wagoner77}, and modernized and made user friendly by Kawano in
1988~\cite{Kawano88}. Nuclear reaction rates were updated in 1993\cite{Smith93}. One
must be careful when making comparisons. First one must consider the numerical
accuracy of the standard code. In 1992 Kawano~\cite{Kawano92} estimated the accuracy
of $Y_P$ to be $6\,\pct$. In 1993, Kernan addressed this issue in more detail and
reported finding a systematic numerical error in the standard
code~\cite{Kernan93,Kernan94}, $\delta Y_P = 0.0017$, large enough to be very
significant at our level of accuracy. Second, the standard code implements certain
physics corrections, namely a correction put in by Wagoner to approximate the Coulomb
correction by scaling all of the weak rates a factor, 0.98, independent of
temperature.

The systematic numerical error discovered by Kernan was measured by comparing the
predictions of the standard code at some (unspecified) integration stepsize to the
predictions as the stepsize became very small; note, however, that the error using
the default stepsize (in Ref.~\cite{Kawano92}) is four times larger. The ``Kernan
correction'' is now routinely added to the results of the standard code. Needless to
say, a simple additive numerical correction is not adequate because other codes
exist; not all users of the standard code use the same stepsize; and the numerical
error can be machine dependent.

\begin{figure}
\centerline{\epsfig{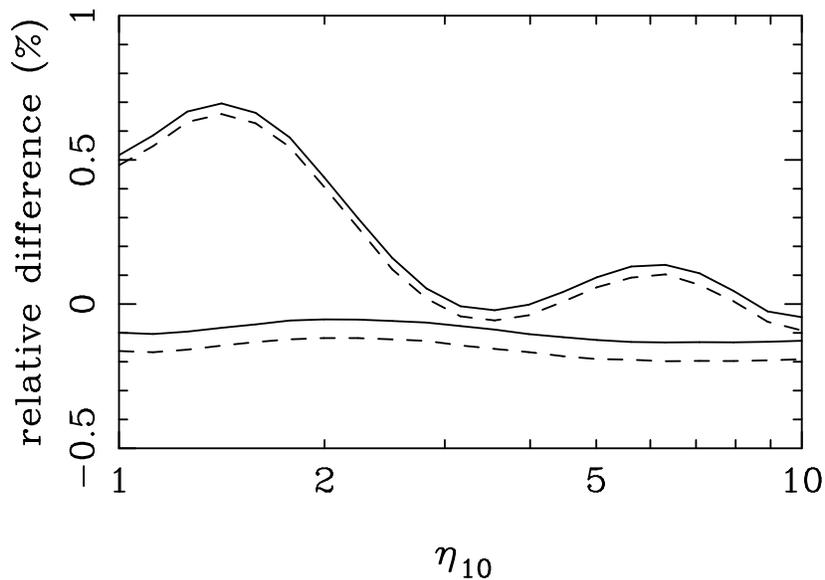}}
\caption{Comparison between the standard code and our code
for \he (lower curves) and D (upper curves). For the solid curves, our very accurate
weak rates were inserted into the standard code. For the dashed curves, the standard
code's weak rate routines were used. (Note: $\eta_{10} \equiv \eta/10^{-10}$).}
\label{fg:bbncomp}
\end{figure}

For our comparisons we took out the Kernan and Coulomb corrections and then made the
stepsizes small enough so that integration errors were negligible. The integration
error for the standard stepsizes (with the two standard stepsize parameters equal to
0.3 and 0.6, respectively) was found to be $\delta Y_P = 0.0073$. With the standard
code configured this way, we compared $Y_P$ and $n_2/n_H$ as a function of $\eta$ in
two scenarios. For the first, we used the standard weak-rate routines to calculate
the weak rates. For the second we used our high-precision weak-rate routines to
calculate the weak rates in the standard code. The results are shown in
Fig.~\ref{fg:bbncomp}. The agreement is excellent: for $Y_P$ the codes differ by less
than 0.15\pct with our weak-rate routines and by less than 0.2\pct with the standard
weak-rate routines. For D the codes agree to better than 0.75\pct.

This agreement gave us confidence that our code calculates $Y_P$ accurately for the
baseline case (without the physics corrections). Of course, the convergence of two
independent codes is not proof that they converge on the correct value. We will
assume that the two codes do indeed converge on the the correct answer, and because
our code was designed, engineered and tested for an error budget, we will use its
results and internal error budget as the baseline for further comparison. The
internal error budget for our code was no greater than 0.1\pct.

\subsection{Nuclear Rate Uncertainties}
\begin{figure}
\centerline{\epsfig{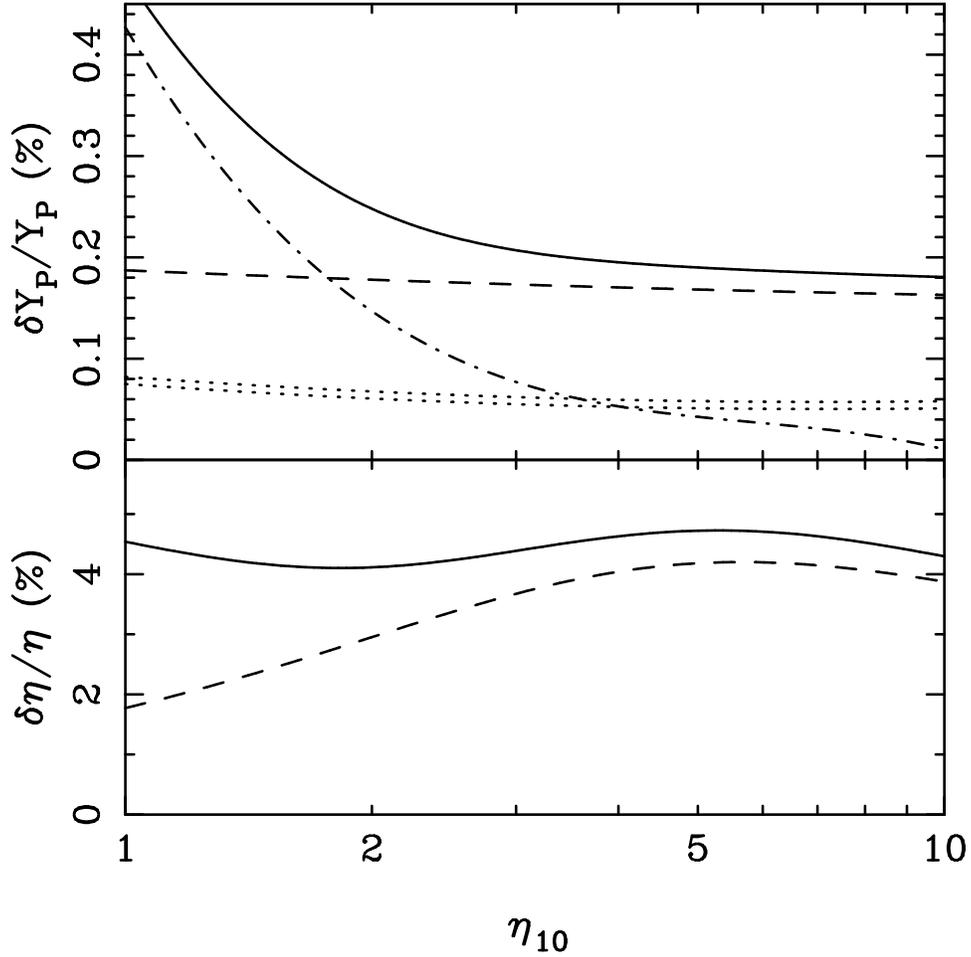}}
\caption{The top panel shows the uncertainty in $Y_P$ due to experimental
uncertainties in nuclear rates, as a function of $\eta$. The solid line shows the
total uncertainty, while the other lines show each nuclear reaction separately. The
dashed line is for $n\leftrightarrow p$, the dashed-dotted line is for
$p(n,\gamma)d$, and the two dotted lines are for $d(d,n)\rm{^3He}$ and
$d(d,p)\rm{T}$. The bottom panel shows the uncertainty in $\eta$ that would result
from the above uncertainties in $Y_P$, when $\eta$ is derived from a perfect
measurement of the $^4$He abundance. The dashed line is for the weak rate
uncertainties alone, while the solid line is for the total nuclear rate
uncertainty. The factor of ten difference in the scales between the two panels is
indicative of the fact that $Y_P$ depends logarithmically upon $\eta$.}
\label{fg:delrk}
\end{figure}

The primordial \he abundance is sensitive to nuclear reactions other than the weak
rates. Several studies of the uncertainties in theoretical abundances due to nuclear
rate uncertainties have been
performed~\cite{Schramm98,Smith93,Kernan94,Krauss90,Krauss94,Hata95}. Here we will
use the results and techniques of the recent work of Fiorentini, et
al~\cite{Fiorentini98}. They use linear error propagation theory to quantify the
effect of experimental uncertainties in the nuclear-reaction rates on the light
element abundance uncertainties and their correlations,
\begin{equation}
\left({\delta Y_P \over Y_P}\right)^2 =  
  \sum_k \lambda^2_k\,\left({\delta R_k\over R_k}\right)^2 \;,
\end{equation}
where the sum $k$ is over nuclear reactions, $\delta R_k$ is the experimental
uncertainty in the rate $R_k$, and $\lambda_k$ is the logarithmic derivative
\begin{equation}
\lambda_k = {\partial \log Y_P \over \partial \log R_k} .
\end{equation}
Fiorentini, et al~\cite{Fiorentini98} calculate the logarithmic derivatives
numerically, using the standard code, and take the experimental rate uncertainties
from Smith, et al~\cite{Smith93}. Contributions to the uncertainty in the \he
abundance arise almost entirely from four rates. Table~\ref{tb:delrk} lists these
rates and their relative experimental uncertainties. Figure~\ref{fg:delrk} shows the
resulting uncertainty in $Y_P$. For $\eta \le 2 \times 10^{-10}$, the reaction
$p(n,\gamma)d$ dominates the error budget. Finally, a recent new analysis of the
experimental uncertainties~\cite{Burles982}, indicates the uncertainties in the
reactions $d(d,n)p$ and $d(d,p)T$ have been overestimated by about a factor of two,
and that the precision of the reaction $p(n,\gamma)d$ could be improved
significantly. Thus it may well be the case that the uncertainty in $\tau_n$
dominates the error budget for all $\eta$. 

\begin{table}
\begin{center}
\begin{tabular}{|c|c|c|c|} \hline
Reaction k & $\delta R_k/R_k $  & $ \delta Y_P / Y_P\;(\eta_{10} = 5.0) $  &
$ \delta Y_P / Y_P\;(\eta_{10} = 1.8)$ \\ 
\hline \hline
$n\leftrightarrow p$ 	& $0.23\pct $ & 0.17\pct & 0.18\pct \\ 
$p(n,\gamma)d$ 		& $7\pct$ & 0.04\pct & 0.17\pct \\
$d(d,n)^3He$ 		& $10\% $ & 0.06\pct & 0.07\pct \\
$d(d,p)T$ 		& $10\% $ & 0.05\pct & 0.06\pct \\
\hline 
\multicolumn{2}{|c|}{Total Uncertainty} & 0.19\pct & 0.27\pct \\
\hline 
\end{tabular}
\end{center}
\label{tb:delrk}
\caption{1-$\sigma$ experimental uncertainties and their effect on $Y_P$. All
nuclear rates whose uncertainties significantly impact $Y_P$ are shown. The weak-rate
uncertainty of 0.23\pct is due to uncertainty in measurements of the neutron mean
lifetime, and assumes that Coulomb, radiative and thermodynamic corrections to the
weak rates are known to better accuracy than this. Note that for $\eta = 5.0\times
10^{-10}$, the neutron mean lifetime dominates the error budget. The bottom row
indicates the RMS total uncertainty in $Y_P$ for these two values of $\eta$. }
\end{table}

\section{Weak Rates}
The primordial \he abundance is very sensitive to the weak rates that maintain the
balance between neutrons and protons. To calculate $Y_P$ to a precision of 0.12\pct
the weak rates must be known to a precision of $0.15\pct$. In addition to numerical
issues discussed earlier, several physical effects are important at this level:
zero-temperature radiative and Coulomb corrections, finite-nucleon mass correction,
and finite-temperature radiative correction.

\begin{figure}
\begin{center}
\epsfig{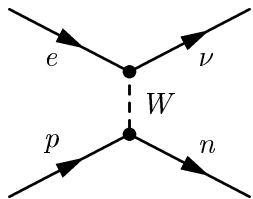}
\end{center}
\caption{Tree level diagram for the process $ep \rightarrow \nu n$.}
\label{fg:borndiag}
\end{figure}

The expressions for the weak rates are derived starting with the tree-level (Born
diagram) shown in Fig.~\ref{fg:borndiag}. For purposes of illustration, we will
consider the process $e^- + p\rightarrow \nu_e + n$.
Without making any approximations the phase space integral for the conversion rate
(per proton) can be simplified to a five-dimensional integral involving the
matrix-element squared $\left|\mathcal{M}\right|^2$~\cite{Lopez97}
\begin{eqnarray}
\Gamma_{ep\leftrightarrow\nu n} 
& = & 
{1 \over 2^9\pi^6 n_p}
\int dp_e dp_p d\cos\theta_p d\cos\theta_\nu d\phi_\nu
\label{eq:fiveD}
\nonumber \\
&   &
\times{p_e^2 p_p^2 E_\nu\over E_e E_p E_n} {1\over {\cal J}} 
\me f_e f_p (1-f_\nu) (1-f_n),  \label{eq:5D} \\
{\cal J} & = & 
1 + { E_\nu \over E_n }
\left( 1 - { ({\bf p}_e+{\bf p}_p) \cdot {\bf p_\nu} \over E_\nu^2 } \right) ,
\label{eq:jacob}
\end{eqnarray}
where $E_e$, $E_p$, $E_\nu$, and $E_n$ denote the energies of the respective
particles and $\cal J$ is the Jacobian introduced in integrating the energy part of
the delta function, and $\left|\mathcal{M}\right|^2$ is summed over initial and final
state spins. The integration limits correspond to the kinematically allowed region in
the five-variable phase space. An expression for $E_\nu = p_\nu$ in terms of the
integration variables $p_e, p_p, \theta_p, \theta_\nu$, and $\phi_\nu$ is given by
\begin{eqnarray}
p_\nu & = & {A^2 B + 2E \sqrt{A^4-m_\nu^2(4E^2-B^2)} \over 4E^2-B^2} ,
\nonumber \\
A^2 & \equiv & 2E_e E_p + 
               m_\nu^2 - m_n^2 - m_e^2 - m_p^2 - 2 p_e p_p \cos\theta_p,
\nonumber \\
B & \equiv & 2\left[ p_e \cos\theta_\nu + p_p\left(\cos\theta_p \cos\theta_\nu
        + \sin\theta_p \sin\theta_\nu \cos\phi_\nu \right) \right] .
\end{eqnarray}
where $E=E_e+E_p$. For more details, see Ref.~\cite{Lopez97}.

\begin{figure}
\begin{center}
\epsfig{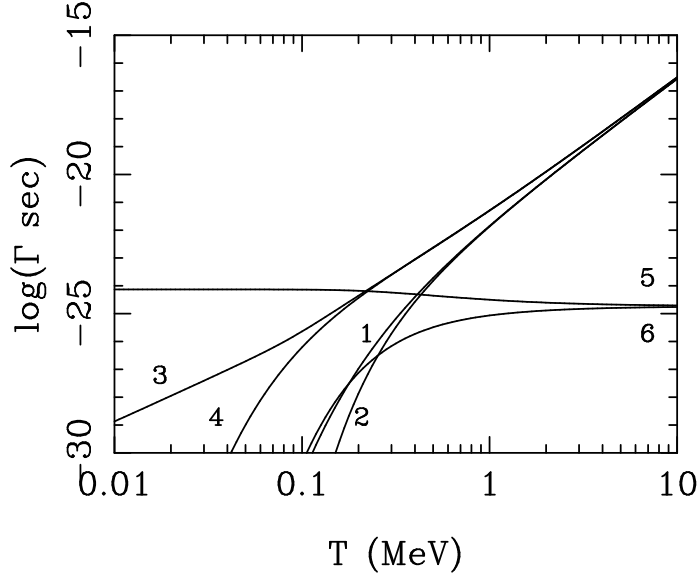}
\end{center}
\caption{Weak rates as a function of temperature (Born diagram, infinite-nucleon-mass
limit): (1) $e p \rightarrow \nu n$, (2) $\nu p \rightarrow e n$, (3) $e n
\rightarrow \nu p$, (4) $\nu n \rightarrow e p$, (5) $n \rightarrow p e
\nu$, (6) $p e \nu \rightarrow n$. Note, freeze-out of the $n/p$ ratio occurs at $T_F
\simeq 0.8\,{\rm MeV}$ and \he synthesis begins at $T \simeq 0.1\,{\rm
MeV}$.}
\label{fg:rates}
\end{figure}

This rate expression is challenging to evaluate for two reasons. First, the
kinematically allowed region in the five-dimensional phase space is not
simple. Second, the full matrix element is complex. Only if the nucleons are assumed
to be infinitely massive, does the expression simplify: $\me \rightarrow 2^5 G_F^2
(1+3g_A^2) E_e E_p E_\nu E_n$. In that limit, the sole kinematical constraint is
$E_p=E_\nu+Q$, ($Q=m_n-m_p=1.293$, MeV), and the rate expression becomes a one
variable integration. Normalizing the rates to the zero-temperature free neutron
decay rate,
\begin{eqnarray}
{1 \over \tau_n} & \equiv \Gamma_{n\rightarrow pe\nu}(T=0) = & {G_F^2(1+3g_A^2)m_e^5
\over2\pi^3} \lambda_0 , \\
\lambda_0 & = & \int_1^q d\epsilon \ \epsilon 
	(\epsilon-q)^2 (\epsilon^2-1)^{1/2} = 1.6333 
\,,
\end{eqnarray}
leads to the well known formula for the process $ep \rightarrow \nu n$:
\begin{equation}
\Gamma_{ep\rightarrow\nu n}^\infty =  
{1\over \tau_n \lambda_0}
\int_q^\infty {\epsilon (\epsilon^2-q^2)^{1/2}\over
\left[1+\exp(\epsilon z)\right]\left[1+\exp((q-\epsilon)z_\nu))\right]} ,
\label{eq:onerate}
\end{equation}
where $T$ is the photon temperature, $T_\nu$ is the neutrino temperature,
$\epsilon \equiv E_e/m_e$, $q\equiv Q/m_e$, $z\equiv m_e/T$, and $z_\nu\equiv
m_e/T_\nu$.  Summing the $n\rightarrow p$ and $p\rightarrow n$ rates yields the
standard weak-rate expressions~\cite{Weinberg72}
\begin{eqnarray}
\Gamma_{n \rightarrow p} & = & { 1 \over \tau_n \lambda_0 }
\left( -\int_{-\infty}^{-1} + \int_1^\infty \right) d\epsilon
{ \eps (\eps-q )^2 \sqrt{\eps^2-1} \over 
  \left(1+e^{-\eps z}\right) \left(1+e^{\left(q-\eps\right) z_\nu} \right) } 
, \nonumber \\
\Gamma_{p \rightarrow n} & = & { 1 \over \tau_n \lambda_0 }
\left( -\int_{-\infty}^{-1} + \int_1^\infty \right) d\epsilon
{ \eps (\eps-q )^2 \sqrt{\eps^2-1} \over 
  \left(1+e^{\eps z}\right) \left(1+e^{\left(\eps-q\right) z_\nu} \right) } .
\end{eqnarray}
The six individual rates are plotted as a function of temperature in
Fig.~\ref{fg:rates}.

\subsection{Zero-Temperature Coulomb and Radiative Corrections}
\begin{figure}
\begin{center}
\epsfig{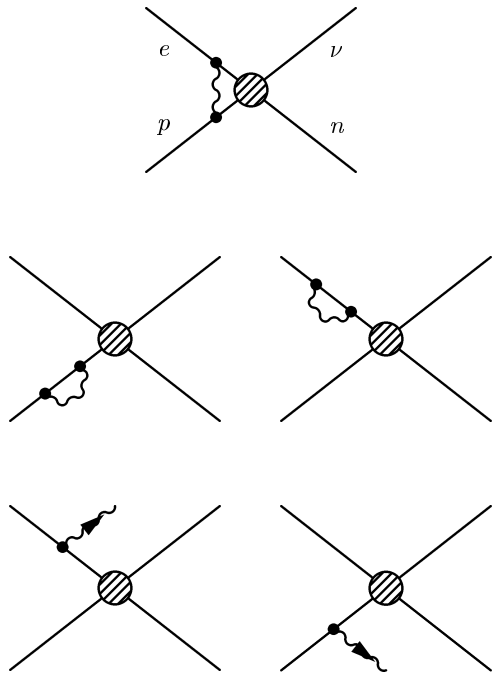}
\end{center}
\caption{Zero-temperature corrections to the process $ep\rightarrow\nu n$. The center 
blob is the charged-current, weak-interaction vertex.}
\label{fg:T0diag}
\end{figure}

To order $\alpha$, the weak rates with zero-temperature Coulomb and radiative
corrections are given by the sum of the interference between the Born diagram
(Fig.~\ref{fg:borndiag}) and the diagrams in Fig.~\ref{fg:T0diag}.

It is conventional to separate the corrections into a Coulomb part proportional to
nuclear charge $Z e$ and a radiative part proportional to $e$. Since $Z=1$ here, this
separation is arbitrary. Dicus et al calculated the Coulomb and zero-temperature
radiative corrections to the weak rates in 1982~\cite{Dicus82}. Summarizing their
results we obtain the following prescription for correcting the rates. First, perform
the zero-temperature radiative corrections by multiplying the integrands of all of
the rates by the factor, 
\begin{equation}
\left[1+{\alpha \over 2 \pi}C(\beta, \rm{y})\right] \,,
\end{equation}
where
\begin{eqnarray}
C(\beta,\rm{y}) & = & 40 + 4(R-1)\left({y \over 3 \epsilon}-{3\over 2}+\ln{2y}\right)
+R\left(2(1+\beta^2)+{y^2\over 6 \epsilon^2}-4\beta R\right) \nonumber \\
&& 
-4\left(2+11\beta+25\beta^2+25\beta^3+30\beta^4+20\beta^5+8\beta^6\right)/(1+\beta)^6
,
\end{eqnarray}
$\beta$ is the electron's velocity and $R = \tanh{\beta}^{-1} / \beta$. Next apply
the Coulomb correction by multiplying the integrand of the rates for $n
\leftrightarrow p e \nu$ and $e p \leftrightarrow \nu n$ by the non-relativistic
Fermi factor,
\begin{equation}
F(\beta) = {2\pi \alpha / \beta \over 1 - e^{-2\pi \alpha/\beta} } .
\end{equation}
The error from using the non-relativistic Fermi function is of order 2\pct of the
Coulomb effect itself~\cite{Wilkinson82}, and so the approximation is fine.  Finally,
$\lambda_0$ must be corrected for Coulomb and zero-temperature radiative effects by
multiplying it's integrand by $\left[1+{\alpha \over 2
\pi}C(\beta,
\rm{y})\right] F(\beta)$. Doing this increases $\lambda_0$ by 7.15\pct, to 1.7501.

\begin{figure}
\centerline{\epsfig{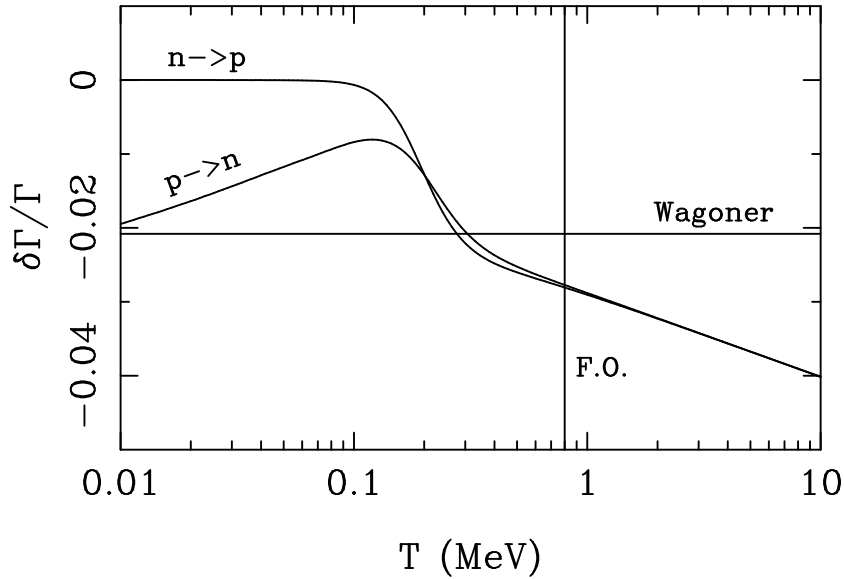}}
\caption{Zero-temperature radiative and Coulomb corrections to the $n\leftrightarrow
p$ rates. The horizontal line is Wagoner's approximation to the Coulomb
correction. The vertical line is at freeze-out.}
\label{fg:T0}
\end{figure}

Figure~\ref{fg:T0} shows the combined zero-temperature corrections. Note that the
corrections are less than or equal to zero for both rates for all temperatures:
decreased weak rates imply earlier $n/p$ freeze-out and an increase in $Y_P$. Our
code calculates the zero-temperature corrections to the weak rates by modifying the
integrands of the rate expressions as described above, and by using the corrected
$\lambda_0$. The zero-temperature corrections yield a change, $\delta Y_P / Y_P =
1.28
\pct$ which is insensitive to the value of $\eta$ over the range $10^{-10} \le \eta \le
10^{-9}$. This result is in agreement with Ref.~\cite{Dicus82}.

Wagoner approximated the Coulomb correction by reducing both the $n \rightarrow n$
and $p \rightarrow n$ rates by 2\pct. This correction, shown by the horizontal line,
is close to the high temperature asymptotic Coulomb correction of
$-2.16$\pct. However, $n/p$ continues to decrease slowly for temperatures lower than
freeze-out, where Wagoner's approximation breaks down. The fact that the real
corrections are less negative in this regime means that the change in $Y_P$ from the
Coulomb correction will be less positive than one would estimate from Wagoner's
approximation. Adding in the zero-temperature radiative corrections brings the total
zero-temperature change in $Y_P$ closer to what would be found using Wagoner's
approximation to the Coulomb correction. Table\ ~\ref{tb:T0} shows $\delta Y_P / Y_P$
for the Coulomb and zero-temperature radiatively separately and summed, compared to
$\delta Y_P / Y_P$ from Wagoner's approximation. Note in particular that the
difference between Wagoner's approximation and the zero-temperature correction is
0.28\pct, which is significant at the 0.1\pct level.

\begin{table}
\begin{center}
\begin{tabular}{|c|c|} \hline
\emph{Correction}	& $\delta Y_P / Y_P$ \\ \hline \hline
Coulomb		 	& 1.04\pct \\
T=0 Radiative		& 0.24\pct \\
Combined		& 1.28\pct \\ \hline
Wagoner's approximation	& 1.56\pct \\ \hline
\end{tabular}
\end{center}
\label{tb:T0}
\caption{Zero-temperature corrections to $Y_P$, compared with change in $Y_P$ from
Wagoner's approximation of the Coulomb correction. These corrections are insensitive
to $\eta$ for $10^{-10}\le\eta\le10^{-9}$.}
\end{table}

\subsection{Finite-Nucleon Mass Correction}
\begin{figure}
\centerline{\epsfig{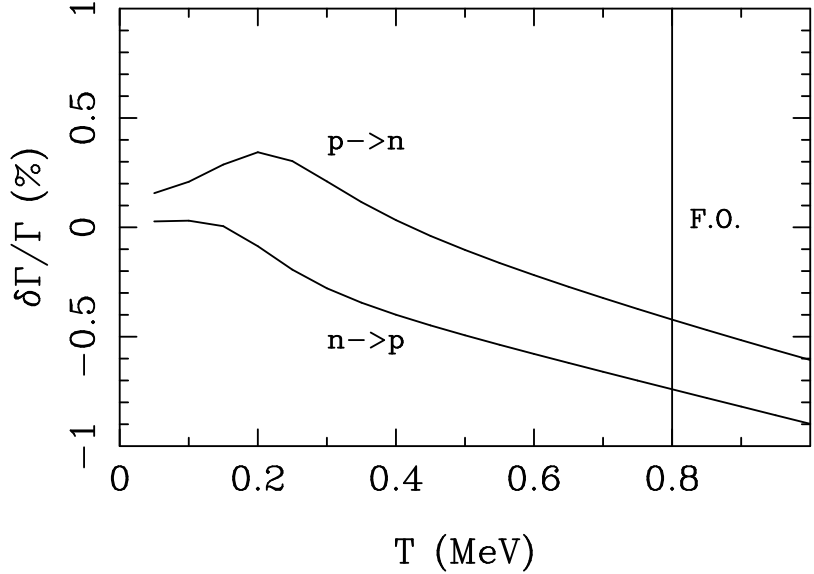}}
\caption{Finite-nucleon-mass correction to the $n\leftrightarrow p$ rates. The
freeze-out temperature, $T_F \simeq 0.8\,\rm{MeV}$, is indicated with a vertical
line.}
\label{fg:finiteM}
\end{figure}

Recall that the standard rate expressions, Eqn.(\ref{eq:onerate}), assume infinitely
massive nucleons. We have calculated the weak rates without this assumption by
numerically integrating the five-dimensional rate integral, Eqn.(\ref{eq:fiveD}),
using the Monte Carlo method~\cite{Lopez97}. Figure~\ref{fg:finiteM} shows the
finite-mass corrections to the $n\leftrightarrow p$ rates. Using the individual rate
corrections we found the corrections to the summed $n\leftrightarrow p$ rates,
\begin{eqnarray}
{\delta\Gamma_{n\rightarrow p} \over \Gamma_{n\rightarrow p}} & \equiv &
{\Gamma_{n\rightarrow p} - \Gamma_{n\rightarrow p}^\infty \over
\Gamma_{n\rightarrow p}^\infty} \\
{\delta\Gamma_{p\rightarrow n} \over \Gamma_{p\rightarrow n}} & \equiv &
{\Gamma_{p\rightarrow n} - \Gamma_{p\rightarrow n}^\infty \over
\Gamma_{p\rightarrow n}^\infty} ,
\end{eqnarray}
where $\Gamma^\infty$ is the rate in the infinite-mass approximation, and $\Gamma$ is
the unapproximated rate. Our corrections are accurate to within a few
percent~\cite{Lopez97}. We incorporated the finite-mass corrections into our code by
modifying the $n\leftrightarrow p$ rates at each temperature by the correction shown
in Fig.~\ref{fg:finiteM}. The resulting correction to $Y_P$ was found to be $\delta
Y_P / Y_P = 0.50\pct$, valid for $10^{-10}\le\eta\le 10^{-9}$.

\subsection{Finite-Temperature Radiative Correction}
\begin{figure}
\centerline{\epsfig{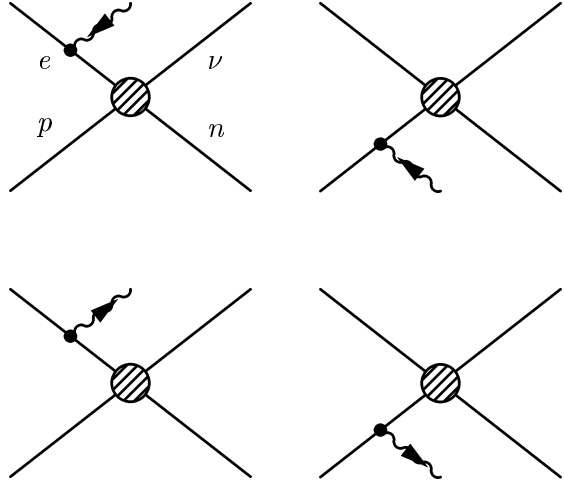}}
\caption{Finite-temperature corrections to the weak rates, i.e., corrections
involving photons from the plasma. The bottom two diagrams represent stimulated
emission.}
\label{fg:real}
\end{figure}

Finite-temperature modifications to the weak rates arise from several sources:
\begin{enumerate}
\item the $(1\pm f)$ quantum statistical factors in the integration over phase space
\label{finT:bose} 
\item a shift in the electron mass \label{finT:me}
\item a change in the neutrino-to-photon temperature ratio \label{finT:Tv}
\item a correction to the photon and fermion propagators \label{finT:prop}
\item the square of the sum of diagrams for processes that involve photons from the
plasma (absorption and stimulated emission); see
Fig.~\ref{fg:real}. \label{finT:real}
\item finite-temperature wave-function renormalization \label{finT:wfr}
\end{enumerate}
Item~\ref{finT:bose} is included in our definition of the Coulomb correction. We
shall define items~\ref{finT:me} and~\ref{finT:Tv} to be part of the thermodynamics
effects, considered later. Therefore, the finite-temperature radiative correction to
the weak rates involves items~\ref{finT:prop}, \ref{finT:real} and \ref{finT:wfr}.

Dicus, et al~\cite{Dicus82}, and Cambier, Primack and Sher~\cite{Cambier82}
calculated the finite-temperature radiative corrections to the weak rates.  Neither
of these papers correctly handle the finite-temperature wave-function
renormalization. In fact, finite-temperature wave-function renormalization is still
an open issue. The difficulty lies in the fact that finite temperature spoils Lorentz
covariance through the existence of a preferred, thermal frame (in this frame the
phase-space distributions are the Bose-Einstein or Fermi-Dirac distributions). The
usual methods for obtaining the wave-function renormalization rely on Lorentz
covariance, so that the appropriate generalization to the finite-temperature case is
not clear.  Donoghue and Holstein~\cite{Donoghue83,Donoghue84} start by assuming a
finite-temperature spinor field -- with creation and annihilation operators obeying
the standard anti-commutation relations -- that satisfies the nonlinear Dirac
equation. They write the propagator in terms of these finite-temperature scalars,
obtaining a finite-temperature wave-function renormalization that is a multiplicative
factor. Sawyer~\cite{Sawyer96}, and Esposito, et al~\cite{Esposito98}, start by
identifying particle states with poles of the propagator, without reference to the
finite-temperature field. They assume that the poles are only perturbatively shifted
from their zero-temperature values. They then identify the finite-temperature
wave-function remormalization with the residue of the propagator at the new pole. The
result is a finite-temperature wave-function renormalization that contains
additional, non multiplicative terms, so that the results of the two alternative
approaches are different (as pointed out by Chapman~\cite{Chapman97}). Furthermore,
the results of the Sawyer differ from Esposito, et al~\cite{Esposito98}, even though
they follow similar approaches. The differences change the rates for some
processes. \emph{However, for the case of the weak rates, the three different
finite-temperature wave-function renormalization results give the same contribution
to the weak rates.}  For convenience, we used the formalism of Sawyer. The correction
to the $en\rightarrow\nu p$ is given as
\begin{eqnarray}
\delta \Gamma & = &
	{e^2 T^4 \over 2^4 \pi^5}G_F^2\left(1+3g_A^2\right) \times \nonumber \\
& &
	\int_0^\infty \int_x^\infty \,du\,dk_v\,
		p_u N_+(u)
		\left[ N_-(k_v) W_\gamma(u,k_v) + 
		 	N_+(v) W_r(u,k_v) \right] ,
\label{eq:finTfirst}
\end{eqnarray}
where $x = m/T$, $p_u = \sqrt{u^2-x^2}$, $v = \sqrt{k_v^2+x^2}$, $N_\pm(u) =
1/(e^u\pm 1)$,
\begin{eqnarray}
W_\gamma(u,k_v) & = &
	\left[ \left( {k_v\over2 p_u} + {u^2\over k_v p_u} \ln{u+p_u \over u-p_u}
	  - {2 u \over k_v} \right) \right]
	\left[ H(u+k_v) + H(u-k_v) - 2H(u) \right] +
\nonumber \\
& &
	\left[ {u\over p_u} \ln{u+p_u\over u-p_u}-2 \right]
	\left[ H(u+k_v) - H(u-k_v) \right] \\
W_r(u,k_v) & = & {k_v H(u)\over 4 p_u v} \left[ 2 u \ln{p_u+k_v\over p_u-k_v} + 
	  v \ln{ m^4 - \left(u v - p_u k_v\right) \over m^4 - 
	    \left( u v + p_u k_v\right)} - {4 k_v p_u u \over p_u^2 - k_v^2}
	\right]
\end{eqnarray}
and 
\begin{eqnarray}
H(w) & = & \nu^2 N(-\nu) \Theta(\nu) , \\
\nu & = & (w+q)
\label{eq:finTlast}
\end{eqnarray}
with $q=Q/T$. The term proportional to $W_r$ is due to finite-temperature wave
function renormalization. To find the correction to the other weak rates, make the
substitutions shown in Table~\ref{tb:finT}.

\begin{table}
\begin{center}
\begin{tabular}{|c|cc|cc|c|c|} \hline
process & lower u-limit & upper u-limit & $e$-Fermi 1 & $e$-Fermi 2 & $\nu$ & $N(\pm
\nu)$ \\ \hline \hline
$e   n\rightarrow \nu p$ & $x$ & $\infty$ & $N(u)$ & $N(v)$ & $w+q$ & $-\nu$ \\
$e   p\rightarrow \nu n$ & $q$ & $\infty$ & $N(u)$ & $N(v)$ & $w-q$ & $-\nu$ \\
$\nu n\rightarrow  e  p$ & $q$ & $\infty$ & $N(-u)$ & $N(-v)$ & $w-q$ & $+\nu$ \\
$\nu p\rightarrow  e  n$ & $x$ & $\infty$ & $N(-u)$ & $N(-v)$ & $w+q$ & $+\nu$ \\
$n \rightarrow p e \nu$ & $x$ & $q$ & $N(-u)$ & $N(-v)$ & $-w+q$ & $-\nu$ \\
$p e \nu \rightarrow n$ & $x$ & $q$ & $N(u)$ & $N(v)$ & $-w+q$ & $+\nu$ \\ \hline
\end{tabular}
\end{center}
\caption{Substitutions in Eqs. (\ref{eq:finTfirst}--\ref{eq:finTlast}) for computing
finite-temperature radiative corrections.}
\label{tb:finT}
\end{table}

\begin{figure}
\centerline{\epsfig{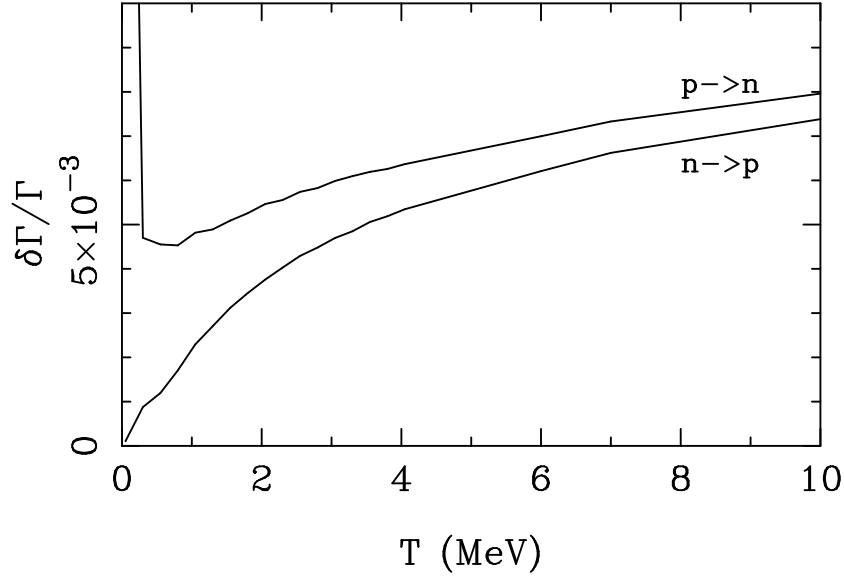}}
\caption{Finite-temperature radiative corrections to the $n\leftrightarrow p$
rates. This plot is to be compared to Fig.~4 in Ref.~[12].}
\label{fg:finiteTR}
\end{figure}

\begin{figure}
\centerline{\epsfig{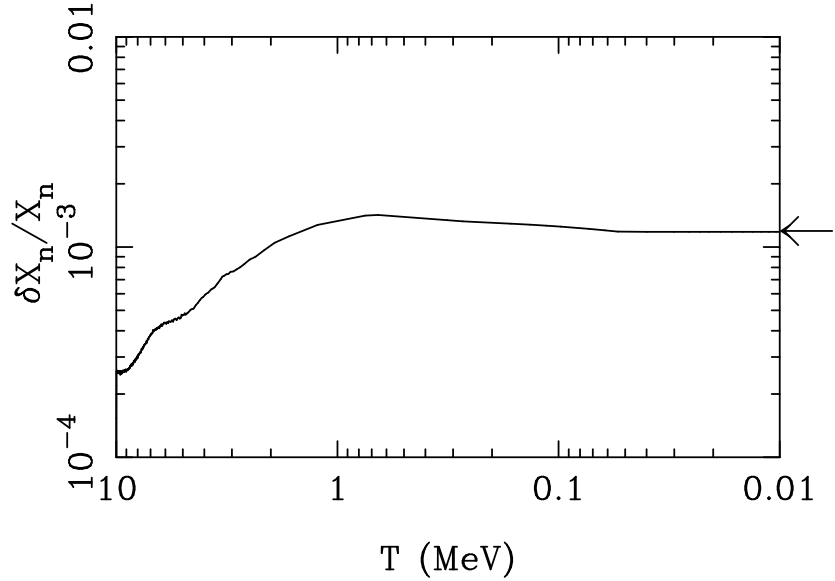}}
\caption{Temperature evolution of the estimated change in neutron fraction $X_n$ due
to finite-temperature radiative corrections. The solid line shows the results of
integrating the perturbation equations; the low-temperature asymptotic solution gives
the correction to $Y_P$, $\delta Y_P/Y_P = \delta X_n/X_n$. The arrow indicates the
final result of substituting the radiative corrections into our full code. The two
methods agree very well.}
\label{fg:delxn}
\end{figure}

We calculated the finite-temperature radiative corrections to each of the weak
rates. The correction to the summed $n\leftrightarrow p$ rates, which match Sawyer's
results, are shown in Fig.~\ref{fg:finiteTR}. The correction formulas are complicated
enough to preclude direct incorporation into our BBN code. Therefore we implemented
these corrections as temperature-dependent fits within the BBN code. The resulting
change in $Y_P$, $\delta Y_P / Y_P = 0.12\pct$, was found to be insensitive to $\eta$
in the range $10^{-10} \le \eta \le 10^{-9}$. Sawyer claims a change of +0.02\pct,
while Chapman claims a change of +0.01\pct. Both Sawyer and Chapman compute the
change in the neutron fraction to estimate $\delta Y_P / Y_P$. To first order in the
perturbation, the equations governing the evolution of the neutron fraction $X_n$ and
its perturbation $\delta X_n$, can be written
\begin{eqnarray}
{d\,X_n \over d\,T} & = & {d\,t\over d\,T}
	\left[-X_n \Gamma_{n\rightarrow p} + \left(1-X_n\right) 
	  \Gamma_{p\rightarrow n} \right] 
\nonumber \\
{d\,\delta X_n \over d\,T} & = & {d\,t\over d\,T} \left\{
	\-\Gamma_{n\rightarrow p} \left( \delta X_n + \gamma_n X_n \right) +
	  \Gamma_{p\rightarrow n} \left[ \gamma_p \left(1-X_n\right) - \delta X_n
	\right] \right\} ,
\end{eqnarray}
where $\gamma_n = \delta\Gamma_{n\rightarrow p} / \Gamma_{n\rightarrow p}$ and
$\gamma_p = \delta\Gamma_{p\rightarrow n} / \Gamma_{p\rightarrow n}$. Then the change
in $Y_P$ is estimated as
\begin{equation}
{\delta Y_P\over Y_P} \simeq \left.{\delta X_n \over X_n}\right|_{onset\:of\:BBN}
\simeq \left.{\delta X_n \over X_n}\right|_{T = 0} .
\end{equation}
In order to have a direct comparison with the results of Sawyer and Chapman, we found
$\delta Y_P / Y_P$ using this method. The evolution of $\delta X_n$ is shown in
Fig.~\ref{fg:delxn}. Our results obtained from this approximation method confirm
those using the BBN code, and differ from Sawyer and Chapman. However, all agree the
change in $Y_P$ is small.

\section{Thermodynamics}
Thermodynamic corrections refer to corrections to the density, pressure and
neutrino-to-photon temperature ratio. There are two effects to consider:
finite-temperature QED corrections to the equation of state of the electromagnetic
plasma, and incomplete neutrino decoupling.

\subsection{Finite-temperature QED Correction}
The finite-temperature QED corrections encompass corrections to the density, neutrino
temperature and electron mass. All of the these corrections follow from the
finite-temperature QED modification to the equation of state of the electromagnetic
plasma. These corrections were calculated by Heckler~\cite{Heckler94} and applied to
cosmology and solar physics. We will follow his approach, correcting a few small
errors.

The \he abundance is sensitive to thermodynamic quantities in several ways. The
energy density determines the expansion rate; changes in the expansion rate affect
the freeze-out temperature, the abundance of free neutrons, and finally $Y_P$. The
next two effects follow from corrections to the electron mass. A change in the
electron mass affects the weak rates directly, and indirectly, by changing the
entropy of the electron-positron plasma at the time neutrinos decouple. Since this
entropy is transferred to the photons when the $e^\pm$ pairs disappear, this changes
the neutrino-to-photon temperature ratio, and affects the weak rates, which are very
sensitive to the neutrino temperature.

\begin{figure}
\centerline{\epsfig{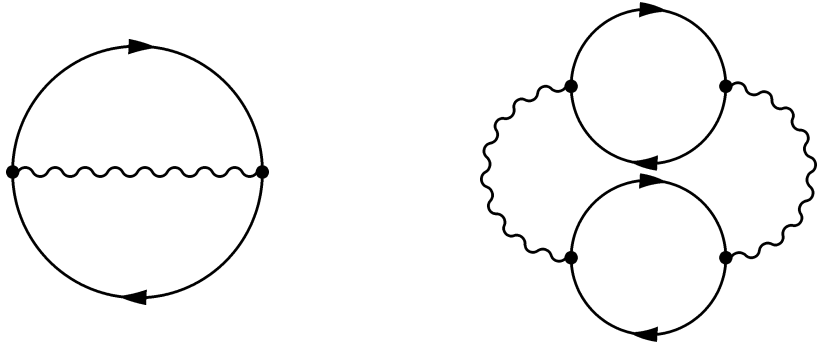}}
\vspace{0.5in}
\label{fg:feyntherm}
\caption{Feynman diagrams that contribute to the correction to the equation of state 
of the electromagnetic plasma. The left diagram produces the order $e^2$ correction,
while the right diagram is the smaller $e^3$ correction.}
\end{figure}

The finite-temperature QED correction to the equation of state can be expressed as a
modification to the pressure of the pressure-weighted, effective number of effective
degrees of freedom, 
\begin{equation}
P(T) = P_0(T) + \delta P(T) ,
\end{equation}
where $\delta P(T)$ is the correction to the pressure and $P_0(T) = (\pi^2/90)\, g_P
T^4$ is the standard expression for the pressure. The change in pressure can be
equated to a change in $g_\rho$, $\delta g_P = 90/(\pi^2 T^4)\,\delta P$. The
correction $\delta P(T)$ can be expressed as an expansion in electron charge $e
\simeq 0.301$: $\delta P(T) = \sum_i \delta P_i(T)$. The Feynman diagrams for the
$e^2$-term and $e^3$-term are shown in Fig.~\ref{fg:feyntherm}. For vanishing
chemical potential the $e^2$ term is~\cite{Kapusta89},
\begin{eqnarray}
\delta P_2(T) & = & 
	-{e^2 T^4\over 6\pi^2} \int_x^\infty du\,{\sqrt{u^2-x^2}\over e^u+1} 
\nonumber \\
& &
	-{e^2 T^4 \over 8\pi^3} \int_x^\infty \int_x^\infty du\,dv\,p_u\,p_v\,N(u)\,N(v)
	\left(4+{x^2\over p_u\,p_v}\ln{u\,v + p_u\,p_v + x^2\over u\,v - p_u\,p_v + x^2}
	\right) ,
\end{eqnarray}
where $x \equiv m_e/T$, $u \equiv E_u / T$, $p_u \equiv \sqrt{u^2-x^2}$ and $N(u) =
1/(1+e^u)$. In the high-temperature limit $T \gg m_e$,
\begin{equation}
\delta P_2(T) \simeq - {5 e^2 T^4 \over 288}
\end{equation}
A similar, but more involved, calculation yields the result for $\delta P_3(T)$ in
the limit $T \gg m$~\cite{Kapusta89},
\begin{equation}
\delta P_3(T) \simeq - {e^3 T^4 \over 36 \sqrt{3} \pi} .
\end{equation} 
At high temperatures, the ratio 
\begin{equation}
{\delta P_2(T) \over \delta P_3(T)} \simeq {1\over e}\;{\sqrt3 \pi \over 2}
\simeq 11 ,
\end{equation}
while both the $e^2$ and the $e^3$-terms are exponentially suppressed for $T \ll
m$. Therefore, to good approximation, we can neglect $\delta P_3(T)$ for all $T$. For
$T\gg m_e, \delta g_\rho = - 25e^2/16\pi^2$. 

\begin{figure}
\centerline{\epsfig{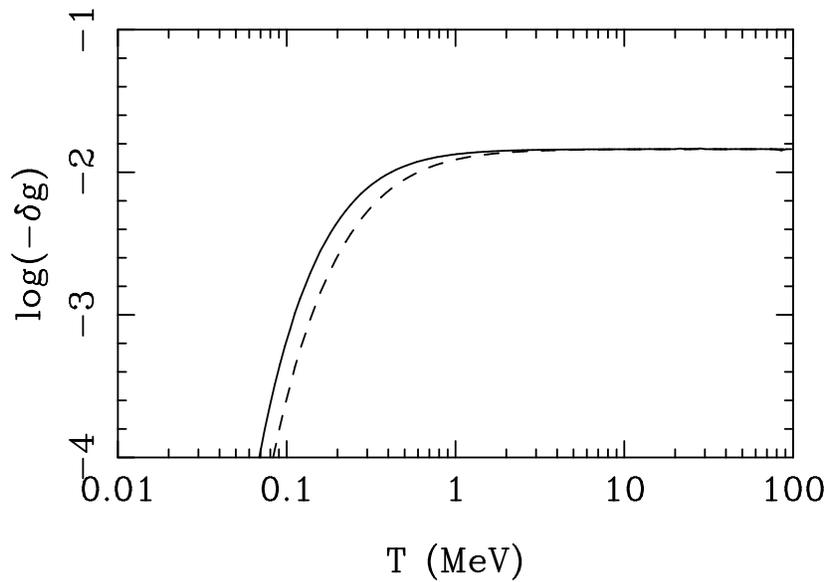}}
\caption{Finite-temperature QED change in pressure-weighted ($g_P$, solid line) and
density-weighted ($g_\rho$, dashed line) relativistic degrees of freedom.}
\label{fg:deltag}
\end{figure}

From the standard thermodynamic relation $\rho = -P + T\,(\partial P/\partial T)$ we
can find the thermodynamic correction to the energy density, $\rho = \rho_0 + \delta
\rho$, where the standard density $\rho_0$ may be written in terms of the
density-weighted effective number of relativistic degrees of freedom, $\rho_0 =
(\pi^2 / 30)\,g_\rho\,T^4$. The change in the density can be written
\begin{equation}
\delta g_\rho = {30 \over \pi^2T^4}
\left(-\delta P + T\,{\partial\over\partial T}\delta P\right)
\overrightarrow{_{T\gg m_e}} -{25\over 16\pi^2} e^2 \,.
\end{equation}
Figure~\ref{fg:deltag} shows $\delta g_\rho$ and $\delta g_P$ as a function of
temperature.

\begin{figure}
\centerline{\epsfig{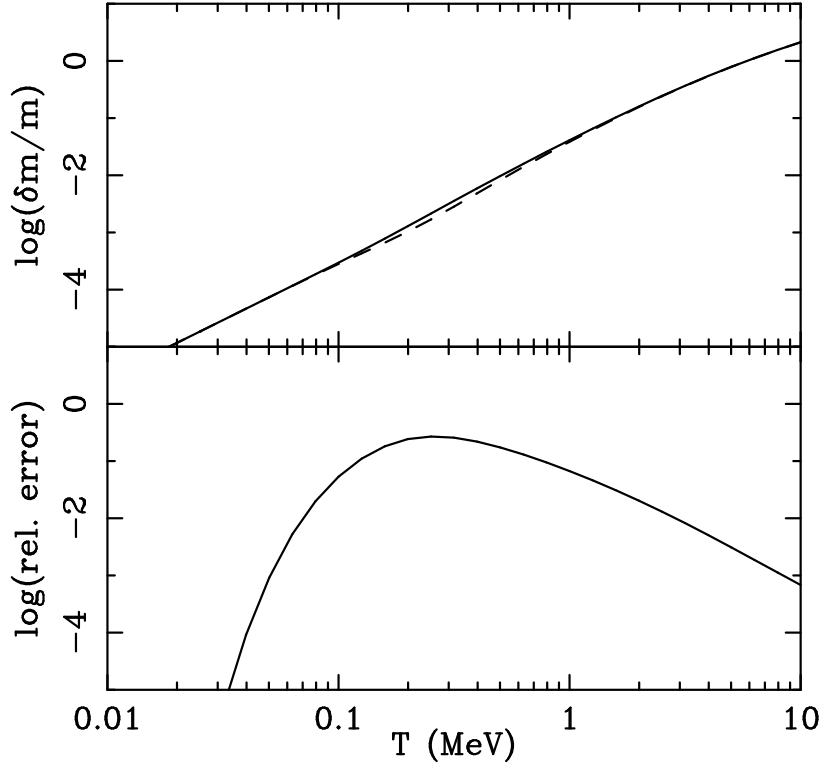}}
\caption{The top panel shows the finite-temperature QED correction to the electron
mass as a function of temperature. The dashed curve neglects the $p$-dependent term,
while the solid curve assumes $p=3\,T$. The bottom panel shows the relative error due
to not including the $p$-dependent term. This error, which is a ten percent
correction to the correction, can be safely neglected.}
\label{fg:deltam}
\end{figure}

\begin{figure}
\centerline{\epsfig{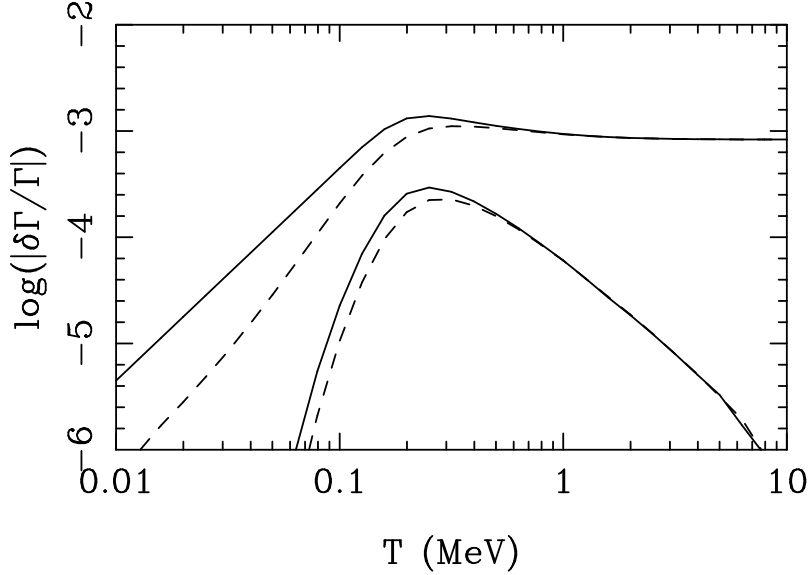}}
\caption{The top curves show the effect of the finite-temperature electron-mass
correction on the weak rates. The solid curve is for $n\rightarrow p$ and the dashed
curve is for $p \rightarrow n$. The bottom curves show the error due to not including
the $p$-dependent term in the mass correction formula.}
\label{fg:delratedelm}
\end{figure}

The finite-temperature QED correction to the pressure is a change in the dispersion
relation of the electrons which can be attributed to a change in the electron mass:
\begin{equation}
E^2 = p^2 + m^2 + \delta m^2 .
\end{equation}
The formula for $\delta m^2$ follows from the definition of the
pressure correction~\cite{Kapusta89}.
\begin{eqnarray}
\delta m^2(p,T) & = & {e^2 T^2 \over 6} + 
	{e^2 T^2 \over \pi^2} \int_x^\infty du\,{k_u\over u}{1\over e^u+1}
\nonumber \\
& & - {e^2 m^2 T \over 2\pi^2 p} \int_x^\infty du\, \ln\left|{p_u+k_u\over
p_u-k_u}\right|{1\over e^u+1} ,
\end{eqnarray}
where $x = m_e/T$, $k_u = \sqrt{u^2 - x^2}$ and $p_u = p/T$.  Figure~\ref{fg:deltam}
shows the finite-temperature QED correction to the electron mass as a function of
temperature. Figure~\ref{fg:delratedelm} shows the effect of the shift in the
electron's mass on the $n\leftrightarrow p$ rates. The lower curves indicate the
error due to not including the momentum-dependent part of the mass correction. For
our calculations, the error is negligible and we neglect the $p$-dependent term in
the mass correction formula.

The final effect of the thermodynamic corrections is a change in the
neutrino-to-photon temperature ratio. This can be derived starting with the
expression for $\delta P(T)$ and tracking the entropy density of the neutrinos and
other particles. Let $s_\nu$ be the entropy density of neutrinos and $s_{\rm{EM}}$ be
the combined entropy density of the electrons, positrons and photons:
\begin{eqnarray}
s_\nu & = & {P_\nu + \rho_\nu \over T_\nu} = {7\pi^2 \over 30} T_\nu^3 \,, \\
s_{\rm{EM}} & = & {P_{e^{\pm}}+\rho_{e^{\pm}}+P_\gamma+\rho_\gamma \over T}
\nonumber \\
& = &
	T^3 \left[{4\pi^2 \over 45} + 
	{2\over 3\pi^2}\int_x^\infty\,du\,{\sqrt{u^2-x^2}\over e^u+1}
	  \left(4u^2-x^2\right) 
	+ {\pi^2\over 90}\left(\delta g_P + 3 \delta g_\rho \right) \right] .
\end{eqnarray}
In the limit that the neutrinos are completely decoupled, the two entropies per
comoving volume are separately conserved: $s_\nu a^3$, $s_{\rm{EM}} a^3 = $ constant,
where $a$ is the scale factor. The small residual coupling of the neutrinos to the
electromagnetic plasma leads to a correction of about $\sim
0.1\pct$~\cite{Dodelson92}, discussed below, which can be ignored here. At high
temperature we have
\begin{equation}
\left.{s_{\rm{EM}} a^3 \over s_\nu a^3}\right|_{T \gg m_e} = {22\over 21} + 
	{1\over 21} \left[\delta g_P(T)+3\delta g_\rho\right(T)] 
	\simeq {22\over21}\left(1-{25\over88}{e^2\over\pi^2}\right) \,,
\label{eq:sratio}
\end{equation}
while for all temperatures,
\begin{equation}
{s_{\rm{EM}} a^3\over s_\nu a^3} = \left({T \over T_\nu}\right)^3 
	\left[ {8\over 21} + 
	{20\over7\pi^4} \int_x^\infty\,du\,{\sqrt{u^2-x^2}\over e^u+1}
	  \left(4u^2-x^2\right) 
	+ {1\over21}\left[\delta g_P(T)+ 3 \delta g_\rho(T) \right]\right] .
\end{equation}
Assuming that the neutrinos decouple at a temperature $T_D \sim 2$ MeV $\gg m_e$ and
taking the ratio of entropies to be given by Eqn.~(\ref{eq:sratio}), it follows that the
ratio of the neutrino-to-photon temperature is
\begin{eqnarray}
\left( {T_\nu\over T} \right)^3 &=&
	{ {4\over11} + {30\over11\pi^4} 
		\int_x^\infty\,du{\sqrt{u^2-x^2}\over e^u+1} \left(4u^2-x^2\right)
	  + {1\over22} \left[ \delta g_P(T) + 3 \delta g_\rho(T) \right] 
	  \over
	  1 - {25 e^2\over 88\pi^2}
	} \,, \\
& \overrightarrow{_{T\ll m_e}} & 
	{4\over11}\left(1+{25e^2\over88\pi^2}\right) 
		\simeq 1.002\left({4\over11}\right) \,.
\end{eqnarray}

\begin{figure}
\centerline{\epsfig{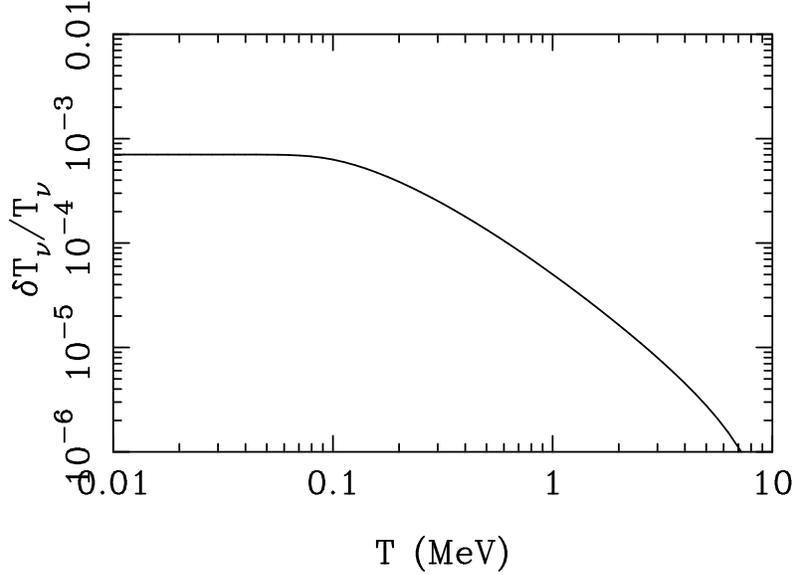}}
\caption{Relative finite-temperature QED change in the neutrino temperature, as a
function of photon temperature. Note that the zero-temperature limit is altered from
the standard value by about $0.08\,\pct$.}
\label{fg:deltv}
\end{figure}

The zero-temperature limit of the neutrino temperature photon temperature relation is
altered~\footnote{This expression differs somewhat from the result obtained by
Heckler~\cite{Heckler94}. He now agrees with our result}. This makes sense
physically: the positive correction to the electron mass means that the
electron-positron plasma has less entropy to give to the photons upon annihilation,
and thus photons are heated less than they would be without the
correction. Figure~\ref{fg:deltv} shows the finite-temperature QED change in neutrino
temperature versus photon temperature.

We incorporated the QED corrections to the equation of state into our code by
changing the energy density, the electron mass in the weak-rate calculations and the
neutrino temperature. The resulting change in $Y_P$, $\delta Y_P / Y_P = + 0.043\pct$
was found to be insensitive to $\eta$ in the range, $10^{-10} \le
\eta \le 10^{-9}$. Dicus, et al~\cite{Dicus82} attempted to calculate the thermodynamic
corrections, and found $\delta Y_P / Y_P = - 0.04\pct$, but only included the effect
of the electron mass on the weak rates. Heckler {\em estimated} the effect on $Y_P$
and found $\delta Y_P / Y_P = + 0.06
\pct$. (It should be noted that his value for the change in neutrino
temperature was incorrect.) In any event, the thermodynamic correction to $Y_P$ is
small.

\subsection{Incomplete Neutrino Decoupling}
The standard code assumes that neutrinos decoupled completely before $e^\pm$
annihilations. It has been pointed out that this assumption is not strictly
valid~\cite{Dicus82}. Neutrinos are ``slightly coupled'' when $e^\pm$ pairs are
annihilated, and hence share somewhat in the heat released. The first calculations
\cite{Dicus82,Herrera89,Rana91} of this effect were ``one-zone'' estimates that
evolved integrated quantities through the process of neutrino decoupling.  More
refined ``multi-zone'' calculations tracked many energy bins, assumed Boltzmann
statistics and made other approximations. \cite{Dodelson92,Dolgov92}. The latest
refinements have included these small effects as well
\cite{Gnedin98,Hannestad95,Dolgov97}. Fields et al~\cite{Fields93} incorporated the
slight effect of the heating of neutrinos by $e^\pm$ annihilations into the standard
code and found a shift in $^4$He production, $\delta Y_P = + 1.5
\times 10^{-4}$, which is insensitive to $\eta$ for $10^{-10} \le \eta \le 10^{-9}$.

\section{Summary}

All of the physics corrections we investigated have been studied elsewhere. However,
not all of them have been implemented in a full code; some have been implemented
incorrectly; and there have been changes in some of the physics corrections. Further,
the issue of numerical accuracy of the standard code has not been comprehensively and
coherently addressed. Finally, the corrections have been implemented in a patchwork
fashion, so that the users of many codes do not know which corrections are in, which
are out, and which may be double counted (e.g., by adding the numerical correction
and running a small stepsize). As noted earlier, the results of a number of BBN codes
gave a 1\pct spread in the prediction for $Y_P$ with the same value of $\eta$ and
$\tau_n$.

The goal of this work was a calculation of the primordial $^4$He abundance to a
precision limited by the uncertainty in the neutron mean lifetime, $\delta\tau_n =
\pm 2\sec$, or $\delta Y_P/Y_P \simeq 0.2\pct$, with reliable estimates of the
theoretical error.  To achieve this goal we created a new BBN code, designed,
engineered and tested to this numerical accuracy.  To this baseline code we added the
microphysics necessary to achieve our accuracy goal -- Coulomb and zero-temperature
radiative corrections, finite-nucleon-mass corrections, finite-temperature radiative
corrections, QED thermodynamical corrections, and the slight heating of neutrinos by
$e^\pm$ annihilations.  These corrections -- coincidentally all positive -- increase
the predicted $^4$He abundance by $\delta Y_P = 0.0049$ or 2\pct.
Table~\ref{tb:results} summarizes these corrections for $\eta = 5 \times
10^{-10}$. For each physical or numerical effect, we have been careful to control the
error in $Y_P$ introduced by approximations or inaccuracies to be well below
0.1\pct. With confidence we can state that the total theoretical uncertainty is less
than 0.1\pct.

Summarizing our work in one number
\begin{equation}
Y_P(\eta = 5\times 10^{-10}) = 0.2462 \pm 0.0004\ ({\rm expt})
\ \ \pm < 0.0002\ ({\rm theory}) .
\end{equation}
Further, the precise value of the baryon density inferred from the Burles-Tytler
determination of the primordial D abundance, $\Omega_B\,h^2 = 0.019\pm
0.001$~\cite{Burles98,Burles98a}, leads to the prediction: $Y_P = 0.2464 \pm
0.0004\,\rm{(expt)}\, \pm 0.0005\,\rm{(D/H)}\,\pm < 0.0002
\,\rm(theory)$.

\begin{table}
\begin{center}
\begin{tabular}{|l|c|cc|cc|} \hline
 & & \multicolumn{2}{|c|}{Cumulative} & \multicolumn{2}{|c|}{Effect Alone} \\ & $Y_P$
 & $ \delta Y_P\,(\times 10^{-4})$ & $\delta Y_P / Y_P\,(\pct)$ & $\delta
 Y_P\,(\times 10^{-4})$ & $ \delta Y_P / Y_P\,(\pct)$
\\ \hline \hline
Baseline & 0.2414 & & & &\\
Coulomb and $T=0$ radiative & 
	0.2445 & +31 & +1.28 &
	+31 & +1.28 \\
finite mass & 
	0.2457  & +43 & +1.78 &
	+12 & +0.50 \\
finite $T$ radiative & 
	0.2460  & +46 & +1.90 &
	+3 & +0.12 \\
QED plasma & 
	0.2461 & +47 & +1.94 &
	+1 & +0.04 \\
residual $\nu$-heating & 
	0.2462 & +49 & +2.00 &
	+1.5 & +0.06 \\ \hline
\end{tabular} 
\end{center}
\caption{Summary of results. For absolute numbers we have picked $\eta = 5.0\times
10^{-10}$. By baseline we mean the results of our BBN code without any of the physics 
effects listed, and with small numerical errors.} 
\label{tb:results} 
\end{table}

Finally, we give two fitting formulae for our high-accuracy $^4$He predictions.  The
first, is accurate to better than 0.05\pct and is valid for $10^{-10} \le \eta \le
10^{-9}$, $N_\nu =3.00$ and $880\sec \le \tau_n \le 890\sec$. In terms of $\zeta
\equiv 10+\log_{10}\eta$,
\begin{eqnarray}
Y_P(\zeta,\tau_n) & = & Y_P(\zeta,885.4\,\rm{sec}) +(\tau_n-885.4\,\rm{sec}) \,\delta
Y_P(\zeta) ,
\nonumber \\
Y_P(\zeta,885.4\,\rm{sec}) & = & \left( a_0 + a_1\,\zeta + a_2\,\zeta^2 +
	a_3\,\zeta^3 + a_4\,\zeta^4 \right) ,
\nonumber \\
\delta Y_P(\zeta) & = & 
	\left( b_0 + b_1\,\zeta + b_2\,\zeta^2 + b_3\,\zeta^3 + b_4\,\zeta^4 \right)
\end{eqnarray}
where the coefficients $a_i,\:b_i$ are given by
\begin{equation}
\begin{array}{ccc}
a_0 = 0.22292 \,, & 
a_1 = 0.05547 \,, & 
a_2 = -0.05639 \,, \\
& a_3 = 0.04587 \,, &
a_4 = -0.001501\\
b_0 = 2.082 \times 10^{-4} \,, &
b_1 = 0.535 \times 10^{-4} \,, &
b_2 = -2.856 \times 10^{-4} \,, \\
& b_3 = 4.672 \times 10^{-4} \,, &
b_4 = 2.420 \times 10^{-4} \,.
\end{array} 
\end{equation}
The second fitting formula is accurate to 0.5\pct and is valid for $10^{-10}\le \eta
\le 10^{-9}$, $880\sec \le \tau_n \le 890\sec$, and $2.5\le N_\nu \le 4.0$.
\begin{equation}
Y_P(\zeta, \tau, N_\nu) = Y_P(\zeta,\tau,3) + \left(N_\nu-3\right) \left( c_0 +
c_1\,\zeta + c_2\,\zeta^2 + c_3\,\zeta^3 + c_4\,\zeta^4 \right) \,,
\end{equation}
where
\begin{equation}
\begin{array}{ccc}
c_0 = 0.01276 \,, & 
c_1 = 0.00409 \,, & 
c_2 = -0.00703 \,, \\
& c_3 = 0.00571 \,, &
c_4 =  -0.00186 \,.
\end{array} 
\end{equation}

\section*{Acknowledgements} 
The BBN code we described in this paper began as a project for a graduate course in
cosmology. R.L. gratefully acknowledges the work of his Red Team colleagues, Moses
Hohman, Mike Joffre, Russel Strickland and Craig Wiegert. Thanks to Gary Steigman for
helpful conversations. This work was supported in part by the DOE (at Chicago and
Fermilab) and by NASA at Fermilab through grant NAG~5--2788, and 5--7092.

\newpage
\bibliography{BBN} 

\end{document}